\documentclass[useAMS,usenatbib,usegraphicx]{mn2e}

\usepackage{fixltx2e}

\def\apj{ApJ}
\def\apjl{ApJL}
\def\aap{A\& A}

\def\araa{ARA\&A}
\def\mnras{MNRAS}

\def\aj{{AJ}}
\def\apjs{ApJS}
\def\bain{{Bull. Astron. Inst. Netherlands}}
\def\mnras{{MNRAS}}
\def\pasj{{PASJ}}

\def\'#1{\ifx#1i{\accent"13\i}\else{\accent"13#1}\fi}
\def\alamenos#1{$^{-#1}$}
\def\ala#1{$^{#1}$}

\def\diezala#1{10$^{#1}$}

\def\be{\begin{equation}}
\def\ee{\end{equation}}

\def\Eg{{E_{\rm grav}}}
\def\e{{\bf e}}

\def\Wext{W_{\rm ext}}
\def\Wextx{W_{{\rm ext,} x}}

\def\Wexty{W_{{\rm ext,} y}}
\def\Wextz{W_{{\rm ext,} z}}
\def\phic{\Phi_{\rm cloud}}
\def\phicor{\Phi_{\rm cor}}
\def\phieff{\Phi_{\rm eff}}
\def\phiext{\Phi_{\rm ext}}
\def\phicentrif{\Phi_{\rm centrif}}
\def\phigal{\Phi_{\rm gal}}

\def\Msun{M_\odot}

\newcommand{\pcc}{\mbox{\,cm}^{-3}}
\newcommand{\pc}{\mbox{\,pc}}

 \title[Gravitational content of molecular clouds and cores] {On
 the gravitational content of molecular clouds and their cores}

 \author[Ballesteros-Paredes, et al.]
{ \parbox{7.0in}{
Javier Ballesteros-Paredes\ala 1
  \thanks{e-mail: {\tt j.ballesteros@crya.unam.mx}},
Gilberto C. G\'omez\ala 1, 
B\'arbara Pichardo\ala 2,
and Enrique V\'azquez-Semadeni\ala 1 \\
} \\ \ala 1 Centro de Radioastronom\'ia y Astrof\'isica,
            Universidad Nacional Aut\'onoma de M\'exico, \\
            Apdo. Postal 72-3 (Xangari), Morelia,
            Michoc\'an 58089, M\'exico \\
     \ala 2 Instituto de Astronom\'ia,
            Universidad Nacional Aut\'onoma de M\'exico, \\
            Apdo. Postal 70-264, 04510, M\'exico, D.F., M\'exico
}

\begin{document}

\date{Submitted to MNRAS, \today}

\pagerange{\pageref{firstpage}--\pageref{lastpage}} \pubyear{2006}

\maketitle

\label{firstpage}

\begin{abstract}

The gravitational term for clouds and cores entering in the virial
theorem is usually assumed to be equal to the gravitational energy,
since the contribution to the gravitational force from the mass
distribution outside the volume of integration is assumed to be
negligible.  Such approximation may not be valid in the presence of an
important external net potential.  In the present work we
analyze the effect of an external gravitational field on the
gravitational budget of a density structure.  Our cases under analysis
are (a) a giant molecular cloud (GMC) with different aspect ratios
embedded within a galactic net potential,
{ including the effects of gravity, shear, and inertial forces,}
and (b) a molecular cloud core embedded within
the gravitational potential of its parent molecular cloud.

{ We find that for roundish GMCs, the tidal tearing due to the shear in
the plane of the galaxy is compensated by the tidal compression in the
$z$ direction.  The influence of the external effective potential on the
total gravitational budget of these clouds is relatively small (up to
$\sim 15-25$\%), although not necessarily negligible.  However, for more
filamentary GMCs, elongated on the plane of the galaxy, the external
effective potential can be dominant and can even overwhelm self-gravity,
regardless of whether its main effect on the cloud is to disrupt it or
compress it.  This may explain the presence of some GMCs with few or no
signs of massive star formation, such as the Taurus or the Maddalena's
clouds.}

In the case of dense cores embedded in their parent molecular cloud,
we found that the gravitational content due to the external field may
be more important than the gravitational energy of the cores
themselves.  This effect works in the same direction as the
gravitational energy, i.e., favoring the collapse of cores.  We
speculate on the implications of these results for star formation
models, { in particular that apparently nearly magnetically critical
cores may actually be supercritical due to the effect of the external
potential.}

\end{abstract}



\section{Introduction}\label{sec:intro}

{All known star formation in the Galaxy occurs within
dense cores in molecular clouds (MCs).  The detailed physical nature
of such molecular clouds and their dense cores has been
a matter of debate over the years.  In particular, the supersonic
linewidths of CO observed in molecular clouds for first time by
\citet{Wilson_etal70} was suggested as indicative of gravitational
contraction by \citet{Goldreich_Kwan74}.  However,
\citet{Zuckerman_Evans74} subsequently argued that, if MCs were
collapsing, the star formation efficiency should be much larger than
observed, and that the molecular gas in our galaxy should already be
exhausted.  These authors suggested, instead, that the supersonic
linewidths were evidence of supersonic turbulence.  This idea was
widely accepted, and turbulence was assumed to be a key ingredient of
molecular cloud support against self-gravity.  For instance,
\citet{deJong_etal80} calculated hydrostatic models of molecular
clouds supported by turbulent (ram) pressure, while \citet{Larson81}
found two scaling relations for atomic and molecular clouds which were
compatible with them being gravitationally bound and in approximate
virial equilibrium.  Since then, many observational studies claim that
molecular clouds and their cores are close to energy equipartition
\citep[most often referred to as ``virial equilibrium'', although this
is not necessarily so; see][]{BP06} between self-gravity, kinetic
energy, and, when the measurements were available, the magnetic energy
\citep[e.g., ][]{MG88a, MG88b, Hei05}.
While some authors supported the idea
that MCs and their cores are self-gravitating \citep[e.g.,][although
with some exceptions, see \cite{BM92}]{MZ92}, others argued that MCs
may not be self-gravitating \citep{Blitz94}, and will ``need'' an
external pressure, in order to be confined \citep[e.g.,
][]{Maloney88}.

With the development of numerical simulations in the last decade, it
became clear that molecular clouds and their cores can exhibit
\citet{Larson81}-type relationships \citep{VBR97, BM02}, and near
energy equipartition \citep{BV95}, but they are not in virial {\it
  equilibrium} \citep{BV97, Shadmehri_etal02, TP04, BP06, Dib_etal07}.
Such clouds and their cores can be also more transient\footnote{By
  transient we mean clouds that either re-disperse into their
  environment, or collapse to form stars,
  { but that in either case do not last much longer than their
  free-fall times.} } than previously thought, a fact that
is compatible with the lack of post-T Tauri stars ($\sim$~10~Myr old
stars) associated to molecular clouds \citep{Briceno_etal97, BHV99,
  HBB01, BH07}.

In most virial balance analyses of molecular clouds and dense
cores, one of the key assumptions is often that the gravitational
term entering the virial theorem for
a cloud or core is given only by its gravitational energy,

\begin{equation}
\Eg = {1\over 2} \int_V \rho \Phi dV,
\end{equation}
where $\Phi$ is the gravitational potential produced by the density
$\rho$ in the volume $V$, thus neglecting the gravitational field due
to the mass distribution outside the volume $V$.  However, in
principle the mass external to the volume of integration can also
influence the energy budget of the cloud/core \citep{Spitzer78, BP06}.
The external distribution of mass, together with external forces
(such as centrifugal and Coriolis forces)
may either tend to tear apart or to
compress the cloud/core under analysis.

The goal of the present work is to numerically evaluate how realistic
is the third Virial Theorem ``Myth'' discussed by \citet{BP06},
namely, that the gravitational term can be approximated by the
gravitational energy.  We aim to compare the gravitational
energy with the effect of external forces for the cases
of giant molecular clouds (GMCs), as well as of the dense
cores within them.  Thus, we evaluate the full external potential
term in the virial theorem for both (i) a typical GMC within the Galaxy,
and (ii) a typical dense core within the gravitational
potential of its parent molecular cloud.

The organization of the paper is as follows.  In \S\ref{sec:tidal} we
provide a physical interpretation for the external gravitational term
$\Wext$, and its contribution to the energy budget of a volume of
interest.  In \S\ref{sec:models} we discuss the cases under analysis,
and in \S\ref{sec:results} we present our results.  Finally, in
\S\ref{sec:discussion} we summarize our results, and discuss their
physical implications, emphasizing their consequences on models of
star formation.

}


\section{Tidal Energy entering the VT}\label{sec:tidal}

The gravitational term entering the VT, given by

\begin{equation}
W \equiv - \int_V x_i\ \rho\ {\partial\Phi \over \partial x_i} \ dV ,
\end{equation}
is usually taken to be equal to the gravitational energy of the cloud,
i.e.,

\begin{equation}
W\simeq \Eg \equiv 1/2 \int_V \rho\ \Phi_{\rm cloud}\ dV,
\label{eq:E_g}
\end{equation}
where $\rho$ is the density, $\Phi$ is the total gravitational
potential, and

\begin{equation}
\Phi_{\rm cloud} = -G \int_V
  \frac{\rho({\mathbf x'})}{|{\mathbf x}-{\mathbf x'}|} dV
\end{equation}
is the gravitational potential of the
cloud, i.e., due only to the mass distribution inside the volume $V$
of the cloud.  This approximation is valid only if the cloud is
isolated, i.e., if the forces from external agents are negligible
compared to the self-gravity of the cloud \citep[e.g., ] []
{Chandra_Fermi53}.  However, MCs in disk galaxies are confined to the
midplane ($z \sim \pm 50$~pc).  Within these disks, most of the mass
in molecular gas is well organized into spiral arms, bars and/or rings
\citep[e.g., ][]{Young_Scoville91, Downes_etal96, Loinard_etal96,
  Loinard_etal99, Dame_etal01, Helfer_etal03}, following the
gravitational potential of the old stars, which dominates the dynamics
\citep{BT87}.  These patterns suggest, thus, that the Galactic
gravitational field may be playing a crucial role in the global
gravitational budget of MCs.  An important issue is thus to determine
the effect of tidal forces on the clouds and their effect on
preventing or inducing their collapse.

\begin{figure}
\includegraphics[width=1.\hsize]{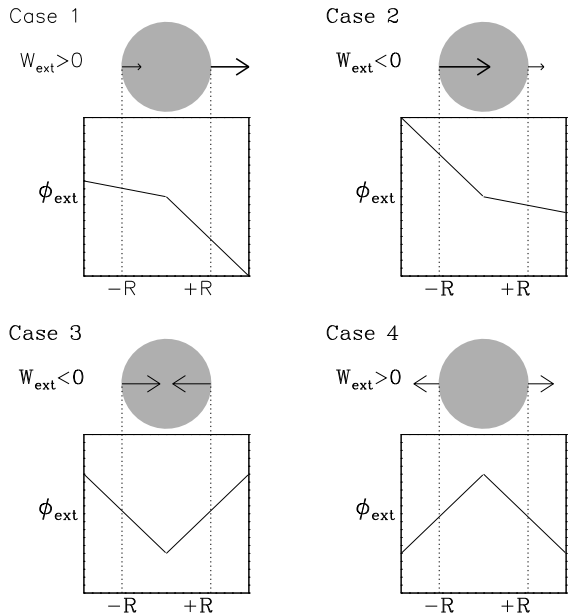}
\caption{\label{fig:ejemplito} The tidal energy, $\Wext$, is positive
if the concavity of the external potential $\phiext$ faces downwards,
and negative if it faces upwards.}
\end{figure}

In order to calculate the contribution of the external potential to
the energy budget of the cloud, let us separate the total
gravitational potential into its component due to the mass of the
cloud alone plus all external agents:

\begin{equation}
\Phi = \Phi_{\rm cloud} + \Phi_{\rm ext},
\label{eq:phi}
\end{equation}
where $ \Phi_{\rm ext}$ is the potential produced by all the mass
outside the volume of the cloud.  Thus, the gravitational term
entering the VT can be written as:

\begin{equation}
W = \Eg + W_{\rm ext}
\label{eq:Wtot}
\end{equation}
where $W_{\rm ext}$ is given by

\begin{equation}
W_{\rm ext}\equiv - \int_V x_i\ \rho\ {\partial \Phi_{\rm ext} \over
\partial x_i} \ dV.
\label{eq:Wext}
\end{equation}
This term represents the contribution to a cloud's gravitational
budget from tidal forces.  Note that we have defined $W_{\rm ext}$
with a minus sign, such that, if it is negative, it contributes to the
collapse, just as the gravitational energy, but if it is positive, it
acts against the gravitational energy by contributing to the
disruption of the cloud.  Furthermore, its sign is given by the mean
concavity of the gravitational potential $\phiext$, being positive if
the concavity faces downwards, and negative if it faces upwards.  To
visualize this, in Fig. \ref{fig:ejemplito} we draw schematically four
different situations in which a spherical cloud is embedded in an
external gravitational potential $\Phi_{\rm ext}$.  In the first case,
(top-left panel)
the forces ($F=-\nabla \phiext$) on the right-hand side of the cloud
are stronger than the forces on the left-hand side.  This is
schematically represented by a darker and larger arrow on the
right-hand side of the cloud, and with a larger absolute value of the
slope of the external gravitational potential.  Since the slope of
$\Phi_{\rm ext}$ is negative, the forces are positive, but stronger at
the right- than at the left-side of the cloud.  The net effect of such
an external field is towards disrupting the cloud.  In fact, for the
left-hand side of the cloud, $x_i < 0$, $\partial \Phi_{\rm
  ext}/\partial x_i < 0$.  The product of these quantities is
positive, and the minus sign of $\Wext$\ (eq. \ref{eq:Wext}) gives a
negative result for $\Wext$.  However, for the right-hand side of the
cloud, $x_i > 0$, and $\partial \Phi_{\rm ext}/\partial x_i < 0$.  The
product is negative, and the minus sign in the definition of $\Wext$
gives a positive value for $\Wext$.  Given the symmetry of the cloud,
and since the forces are larger on the right-hand side, the total
value of $\Wext$\ is positive, i.e., the external field acts against
the gravitational energy.  In other words, the cloud is being torn
apart by the external gravitational field.

For the second case (top-right panel), the
dominant contribution is that from the left-hand side, and the cloud
suffers a net external compression.  In this case, a similar analysis
gives $\Wext<0$.
Cases 3 and 4 (respectively, bottom left and right panels)
can be analyzed in a similar way, and
$\Wext$\ is negative and positive, respectively.  In summary,
$\Wext$\ {\it contains the net effect of the tidal forces over the
  whole volume of the cloud,} and {\it its sign is defined by the
  curvature of the gravitational potential field.}

Before ending the present section, it is useful to define 
\begin{eqnarray}
\Wextx &\equiv & - \int_V\ x\ \rho\ {\partial\phiext / \partial x}\ dV ,
\nonumber\\ 
\Wexty &\equiv & - \int_V\ y\ \rho\ {\partial\phiext / \partial y}\ dV ,
\label{eq:definicionesWext} \\
\Wextz &\equiv & - \int_V\ z\ \rho\ {\partial\phiext/\partial z}\ \nonumber dV ,
\end{eqnarray}
%
such that $\Wext = \Wextx + \Wexty + \Wextz$, in order to understand
how the three-dimensional potential works in each direction.


\section{Modeling the gravitational field and the volume of
    integration}\label{sec:models}

We evaluate the virial gravitational term in two main cases: a giant
molecular cloud (GMC) within a disk galaxy, and a molecular cloud core
at different positions within its parent molecular cloud.  In every
case, following eq. (\ref{eq:phi}), we will assume that the total
gravitational potential $\Phi$ is given by the addition of two
independent potentials: that produced by the mass of the object under
analysis (GMC or core, which will be called $\phic$), and the external
potential (from the Galaxy or the GMC), which we will call $\phiext$.


\subsection{Giant Molecular clouds within a disk galaxy}

We first consider the effects of a galactic gravitational potential
field over a giant molecular cloud.  We consider a set of prolate
clouds with constant density $n=50$~cm\alamenos 3, and major axis of
25~pc.  The aspect ratios for each case are: 10:10 (spherical case),
10:8, 10:6, 10:4, 10:2, and 10:1.  {We align those spheroids in
  both the $x$- and $y$-directions, in order to see how the cloud's
  energy budget varies as a function of the relative orientation of the
  spheroids and the spiral arms within the galaxy.}

{ 

The clouds are located within a spiral galaxy with a gravitational
potential field given by the model implemented by
\citet{Pichardo_etal03}.  This model uses an axisymmetric background
potential that assembles a bulge, a flattened disk with a scale-height
of 250 pc, as proposed by \cite{Miyamoto_Nagai75}, and a massive halo
extending to a radius of 100 kpc, as proposed by
\citet{Allen_Santillan91}.
The main adopted parameters are $R_0 = 8.5$ kpc as the Sun's
galactocentric distance, and $V_0(R_0) = 220$ km s$^{-1}$ as the
circular velocity at the Sun's position. The total mass is $9\times
10^{11} M_{\odot}$, and the local escape velocity is 536 km
s$^{-1}$. The local total mass density is $\rho_0 = 0.15 M_\odot$
pc$^{-3}$.  The resulting values for Oort's constants are $A = 12.95$
km s$^{-1}$ kpc$^{-1}$ and $B = -12.93$ km s$^{-1}$ kpc$^{-1}$.  The
full expressions of the axisymmetric potential can be found in
\citet{Allen_Santillan91}.

}

In addition to the axisymmetric potential, we have
implemented a bi-symmetric spiral model constructed
with oblate spheroids as those proposed by
\citet{Schmidt56}
using the spiral logarithmic locus proposed by
\citet{Roberts_Huntley_vanAlbada79}.  The parameters for the spiral
arms based on self-consistency studies and a compilation of
observational results \citep{Pichardo_etal03} are: the pitch angle is
$i_p=15.5^o$; the galactocentric radii at which the arms start and end
are, respectively $r_s=3.3$ kpc and $r_e=12$ kpc; the width of the
spiral arms is 2 kpc.  The density fall along the spiral arm is
exponential with the same scale-length as the disk (2.5 kpc). Finally,
the ratio of spiral mass to disk mass is $M_S/M_D=0.0175$, which
represents a lower limit on the range given by self-consistency
analysis.

Every point in the volume of integration will be analyzed in a frame
of reference located at the center of the cloud, which rotates around
the Galactic center.  This adds a centrifugal potential that produces
the force

\begin{equation}
F_{\rm centrif} =
- {\partial \phicentrif \over \partial r} = \Omega_0^2 r \ {\hat \e}_r,
\label{eq:centrifugal}
\end{equation}

\noindent
where $\Omega_0$ is the angular velocity of the
reference frame that rotates with the center of the cloud,
$r$ is the galactocentric distance of the fluid element under
consideration, and ${\hat \e}_r$ is the unit vector in the direction
of the galactocentric radius.  It is also necessary to account for the
Coriolis term in our modified potential:

\begin{equation}
F_{\rm cor} =
- {\partial \phicor \over \partial r} = 2 \Omega_0 v \ {\hat \e}_r ,
\label{eq:coriolis}
\end{equation}
{ where $v$ is the velocity of the point in the volume of
  integration, measured in the frame of reference of the center of the
  cloud, i.e., in the frame that moves with angular frequency
  $\Omega_0$ } With this in mind, the effective potential entering in
  eqs. (\ref{eq:Wtot}) and (\ref{eq:Wext}) is given by

\begin{eqnarray}
\nabla\phieff &=& \nabla \phigal + \nabla\phicentrif + \nabla \phicor
  \nonumber \\ 
            &=& \nabla \phigal - ( \Omega_0^2 r \ +\ 2 \Omega_0 v ) \
  {\hat \e}_r .
 \label{eq:pot_effectivo}
\end{eqnarray}

{ Note that this potential assumes perfect circular orbits, while
  actual orbits are not necessarily perfect circles.  However, the
  velocities in both cases are similar within a factor not larger than
  10\%.  Since the velocities entering in eq. (\ref{eq:pot_effectivo})
  affect two of the three terms, we do not expect changes larger than
  10\%. }


\subsection{Dense cores within a molecular cloud}
\label{sec:core_setup}

In addition to the molecular cloud embedded within a spiral galaxy, we
analyze the gravitational budget of smaller structures: dense cores
embedded within its parent GMC.  Although molecular clouds and their
cores exhibit substantial substructure over a wide range of sizes
\citep[e.g., ][]{Falgarone_etal91}, we will consider, for simplicity,
that our test core is spherical, with constant density
($n=3500$~cm\alamenos 3), and size $R_{\rm core}=0.5$~pc, which are
typical of dense dark cloud cores \citep{Troland_Crutcher08}.  We also
model the gravitational potential field of the parent molecular cloud
with a \citet{Plummer1911} potential, as used by
\citet{Gieles_etal06}, who analyzed the evolution of a stellar cluster
interacting with a giant molecular cloud.  Such potential is written
as

\begin{equation}
  \phiext = -\frac{  \mathrm{G} M_{\rm GMC} }{\sqrt{ r^2 + a^2 }},
\end{equation}

\noindent
with $a = r_{\rm GMC} / 2$.  Following \citet{Gieles_etal06}, we
relate the mass and size of the GMC by a \citet{Larson81} type
relation\footnote{Note that this relation has been highly
  questioned, since it seems to be more a consequence of observational
  biases, rather than a true relationship \citep{Kegel89, Scalo90,
    VBR97, BM02}.  However, for the sake of simplicity, the models of
  clouds that we analyzed follow this relationship.}.

\begin{equation}
  M_{\rm GMC} = 540 M_\odot \left(\frac{r_{\rm GMC}}{1\pc}\right)^2,
  \label{eq:r_gmc}
\end{equation}

\noindent
which, for $M_{\rm GMC}=10^4 M_\odot$ yields $r_{\rm GMC} = 4.30\pc$.
{ For these parameters, the GMC's central density is
$4.13\times 10$\ala 3\,cm\alamenos 3.}

Since GMCs frequently exhibit elongated or filamentary structure, we
 consider two other simple cases of modified Plummer spheres (a
 detailed analysis including a variety of more realistic parent cloud
 structures will be presented in a future contribution). In
 the first place, we studied the superposition of two Plummer
 spheroids: one centered at $x=0$ and with mass $M_{\rm GMC}$, and the
 second centered at $x=3 r_{\rm GMC}$, with a mass $4 M_{\rm GMC}$, and
 a radius twice as large, according to eq. (\ref{eq:r_gmc}).

Finally, we modify the \citet{Plummer1911} potential to obtain an
elongated structure, in order to mimic a filamentary cloud.  In this
case, the external potential is given by

\begin{equation}
  \phiext = -\frac{\mathrm{G} M_{\rm GMC} }{\sqrt{ \tilde{r}^2 + a^2 }},
\label{eq:plummer_esferoid}
\end{equation}

\noindent
where

\begin{equation}
  \frac{\tilde{r}(x,y,z)}{r_{\rm GMC}} =
                 \left[ \left( \frac{x}{r_x} \right)^2
                      + \left( \frac{y}{r_y} \right)^2
                      + \left( \frac{z}{r_z} \right)^2 \right]^{1/2},
\label{eq:plummer_esferoid_radius}
\end{equation}

\noindent
$r_x=r_{\rm GMC} f^{ 2/3}$ and $r_y=r_z=r_{\rm GMC} f^{-1/3}$, where
$f$ is the aspect ratio of the resulting spheroid.  For $f=10$,
$r_x=20\pc$ and $r_y=r_z=2\pc$.  The parameters $M_{\rm GMC}$ and
$r_{\rm GMC}$ are related also by eq. (\ref{eq:r_gmc}).


\section{Results}\label{sec:results}

\subsection{The molecular cloud within a disk galaxy}

\begin{figure*}
\includegraphics[width=0.95\hsize]{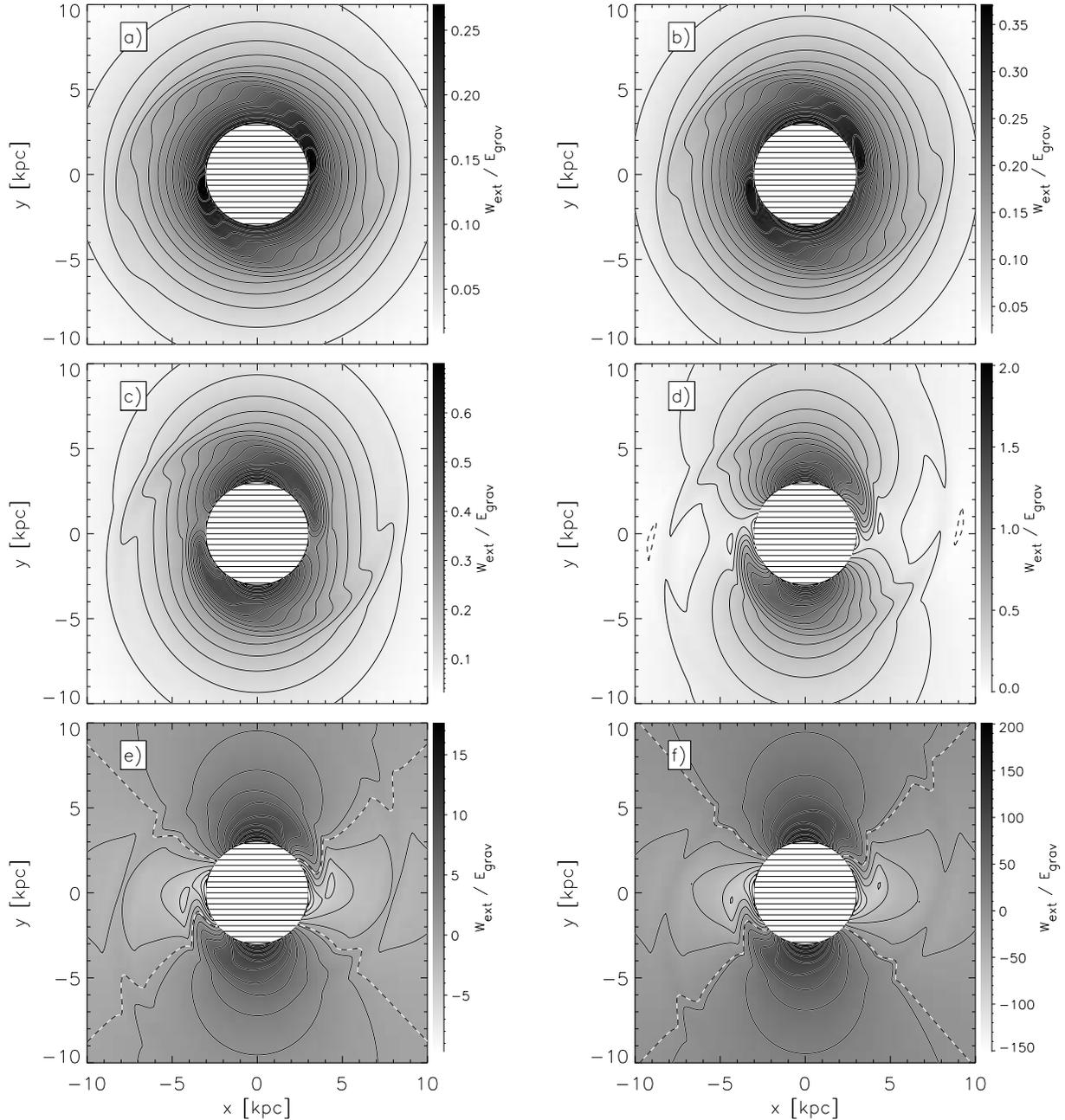}
  \caption{ Grayscale map of the ratio $\Wext/\Eg$ for GMCs
    (represented as prolate spheroids) aligned with the $x$ axis.  A
    central circle ($r\le 3$~kpc) is excluded from the calculation
    since there are some missing physical ingredients, such as a
    galactic bar potential.  The dashed line represents the $\Wext=0$
    level.  Each case is a prolate spheroid, with larger-to-smaller
    ratio of (a) 10:10 , (b) 10:8, (c) 10:6, (d) 10:4, (e) 10:2 and
    (f) 10:1, respectively.  Note that, for clouds with small axis
    larger than 10~pc (cases a, b, c, and d), the ratio is positive,
    i.e., $\Wext$ acts in the same sense as gravity.  This is due
    mainly to the strong curvature of the gravitational potential
    along the galactic height $z$, which dominates in compressing the
    cloud against any tidal disruption imposed by the shear and/or
    tidal streams from the spiral arms. However, for clouds elongated
    in the plane of the galaxy by a factor of $\sim$~10:4 or more
    (cases e and f), either tidal stretching or compression can
    dominate the energy budget, depending on the position and
    orientation of the filament in the galaxy.  This result suggests
    that the intensity of star formation of a cloud may have a strong
    dependence on its morphology, its position and its orientation
    within the galaxy. }
  \label{fig:Wext_x}
\end{figure*}

\begin{figure*}
\includegraphics[width=0.95\hsize]{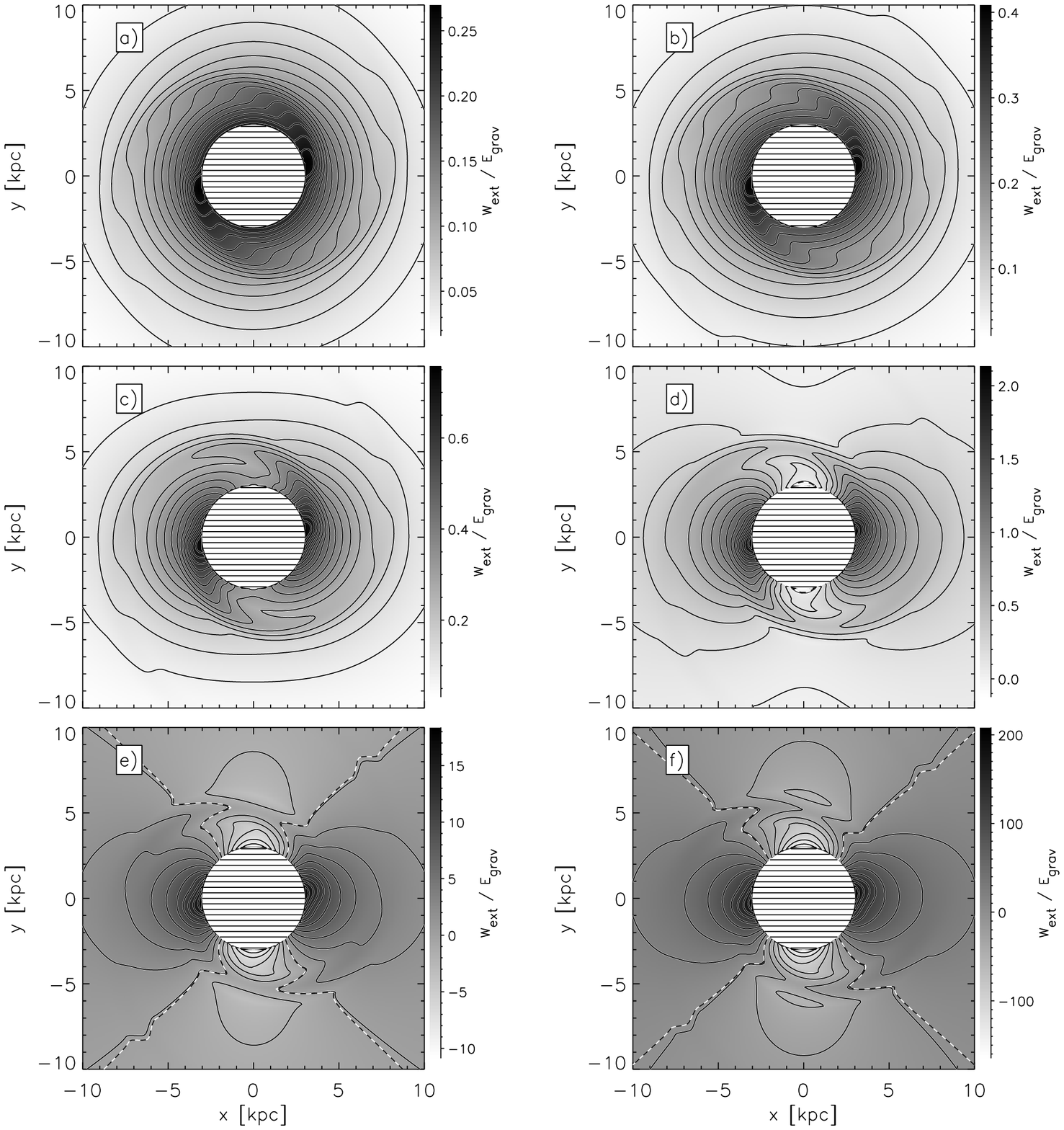}
  \caption{Same as Fig.~\ref{fig:Wext_x}, but for spheroids aligned
    with the $y$-axis.}
  \label{fig:Wext_y}
\end{figure*}

Figures~\ref{fig:Wext_x} { and \ref{fig:Wext_y}} show the ratios
$\Wext/\Eg$ through the galaxy at $z=0$ for our prolate spheroidal
GMCs ($n=50\pcc$, long axis $R_{\rm cloud}=25$~pc on the plane of the
galaxy, and aspect ratios of 10:10, 10:8, 10:6, 10:4, 10:2 and 10:1,
along the $x$- and { $y$-directions, respectively}).

{
The first important result is that the term due to the external
potential, $\Wext$, does not seem to be negligible in most of places
within the galaxy.  This is in clear contrast with the typical
assumptions in the literature, where the gravitational energy $\Eg$ is
thought to be the quantity to be compared to other kind of energies in
order to explain collapse, support, or expansion.  Secondly, the
maximum and minimum possible values of the ratio $\Wext/\Eg$ are quite
similar for both ($x$- and $y$- alignment ) cases, although the
structure of the images differs substantially, specially in the cases
with large aspect ratios (panels d,e,f in both figures).

We also note that, for the more roundish clouds (cases a, b, c, d),
the effective potential of the galaxy works in the same direction as
the self-gravity, giving positive values of the term $\Wext/\Eg$.
This is so because, while the effective potential imposes a shear that
tries to disrupt the clouds in the plane of the galaxy (i.e., $\Wextx
+ \Wexty >0$), the gravitational potential of the galaxy along the $z$
axis has a confining effect, and it may dominate if the cloud is long
enough along $z$ compared to its dimensions on the plane of the
galaxy.  The situation is, however, somewhat different for more
elongated clouds that do not extend very much in $z$ (panels d e, f).
In these cases, the effective gravitational field of the rotating
cloud may work either by compressing (darker zones) or disrupting
(light zones) the cloud, depending on the region where the cloud is
located within the galaxy.  Furthermore, by comparing
Figs.~\ref{fig:Wext_x} { and \ref{fig:Wext_y}}, it can be noticed
that if the filamentary cloud is perpendicular to the radius of the
galaxy, the external field works toward compressing the cloud (darker
zones), while if it is parallel to the radius of the galaxy, the
external field tries to disrupt it (light zones), even if it is close
to the spiral arms.  This result is due to the differential rotation,
since a cloud aligned with the radius of the galaxy will feel the
galactic differential rotation much more stronger than a cloud that is
perpendicular to it.


Finally, it is interesting to note that, for the roundish cases, the
external contribution to the gravitational budget is not larger than
$\sim 25$\%.  However, for the more elongated cases, this contribution
can be a large factor, suggesting that some filamentary GMCs may
actually be gravitationally unbound.  This could be the case, for
instance, of those few GMCs that
exhibit no signs of massive star formation, yet have masses of the
order of \diezala 4-\diezala 5$~\Msun$ (e.g., Taurus, G~216-2.5,
Coalsack).

One may ask how much these results can vary with the parameters of the
cloud (size and density).  Let us assume that the cloud is ``small
enough'', so that the external forces over the cloud can be
approximated by linear functions of the distance to the center of the
cloud, then $\nabla \phiext \propto R_{\rm cloud}$.  In this case,
$\Wext \propto \rho R_{\rm cloud}^5$, while $\Eg \propto M^2/R_{\rm
cloud} \propto \rho^2 R_{\rm cloud}^5$. Then, the ratio $\Wext / \Eg
\propto 1/\rho_{\rm cloud}$.  Thus, for a given shape of the cloud,
the ratio $\Wext / \Eg $ can be scaled to different densities as
$1/\rho_{\rm cloud}$.  In our case, ``small enough'' is smaller than
$\sim$ 50~pc.

In the present work, we have decided to keep the density constant and
vary the shape of the clouds. However, it is possible in principle
that the resulting increase of $\Wext / \Eg $ for spherical clouds for
elongated clouds could be due to a decrease in the clouds' mass, and
therefore their gravitational energy as the shape becomes more
elongated, giving the impression of an increase in the relative
importance of $\Wext$.  Therefore, another plausible scenario to
explore would be to fix the mass, or the column density, while varying
other parameters of the cloud.  However, our results can be translated
to those cases.  For instance, since $\Wext / \Eg \propto 1/\rho_{\rm
cloud}$, by keeping constant the mass of the clouds, the ratio $\Wext
/ \Eg $ grows with size as $R_{\rm cloud}^3$.  In the case of constant
column density, the $\Wext / \Eg \propto R_{\rm cloud}$.  In order to
show that this is the case, in Fig.~\ref{fig:variaciones} displays
three panels with the ratio $\Wext/\Eg$ for spherical clouds as a
function of $x\geq 0$ $(y=0, z=0)$ for clouds with different densities
but same size (left panel), different sizes but same mass (middle
panel), and different sizes, but same column density (right panel).
It can be seen that $\Wext/\Eg \propto 1/\rho$ for all cases (and then
$\propto R^3$ and $\propto R$ for the constant mass and column density
respectively).  Thus, for a given cloud, one can calculate its mass,
or its column density, and then, calculate how much the ratio
$\Wext/\Eg$ will vary if the size varies too.

\begin{figure*}
\includegraphics[width=1\hsize]{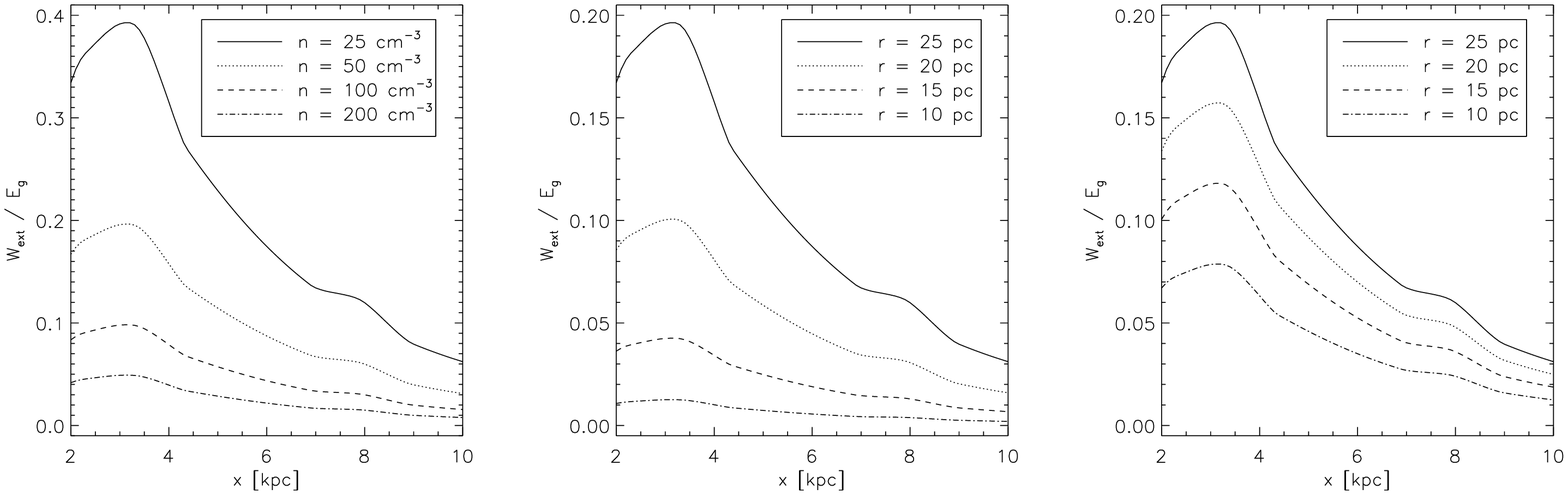}
  \caption{ Variation of the ratio $\Wext/\Eg$ for spherical clouds
    of the same size but different densities (left panel), same mass
    but different sizes (middle panel), and same column density but
    different sizes (right panel).  Note that $\Wext/\Eg \propto
    1/\rho$, for all cases, which implies that $\Wext/\Eg \propto
    R_{\rm cloud}^3$ and $\Wext/\Eg \propto R_{\rm cloud}$
    for the cases of constant
    mass and column density, respectively. }
  \label{fig:variaciones}
\end{figure*}

We finally note that our results do not change strongly by placing the
clouds above or below the plane of the galaxy.  At first glance, one
might think that a cloud located 50 or 100~pc away from the midplane
will experience a tidal stress along $z$, rather than a tidal
compression.  However, we note that the gravitational potential along
$z$ has always positive concavity, and then the situation is just like
the one shown in Fig~\ref{fig:ejemplito}, cases 2 or 3.  It is
important to note, however, that the radius of curvature of the
gravitational potential grows with $z$, and thus the vertical
compression decreases as $z$ increases. 

}


\subsection{The core within a GMC}

\begin{figure}
  \includegraphics[width=84mm]{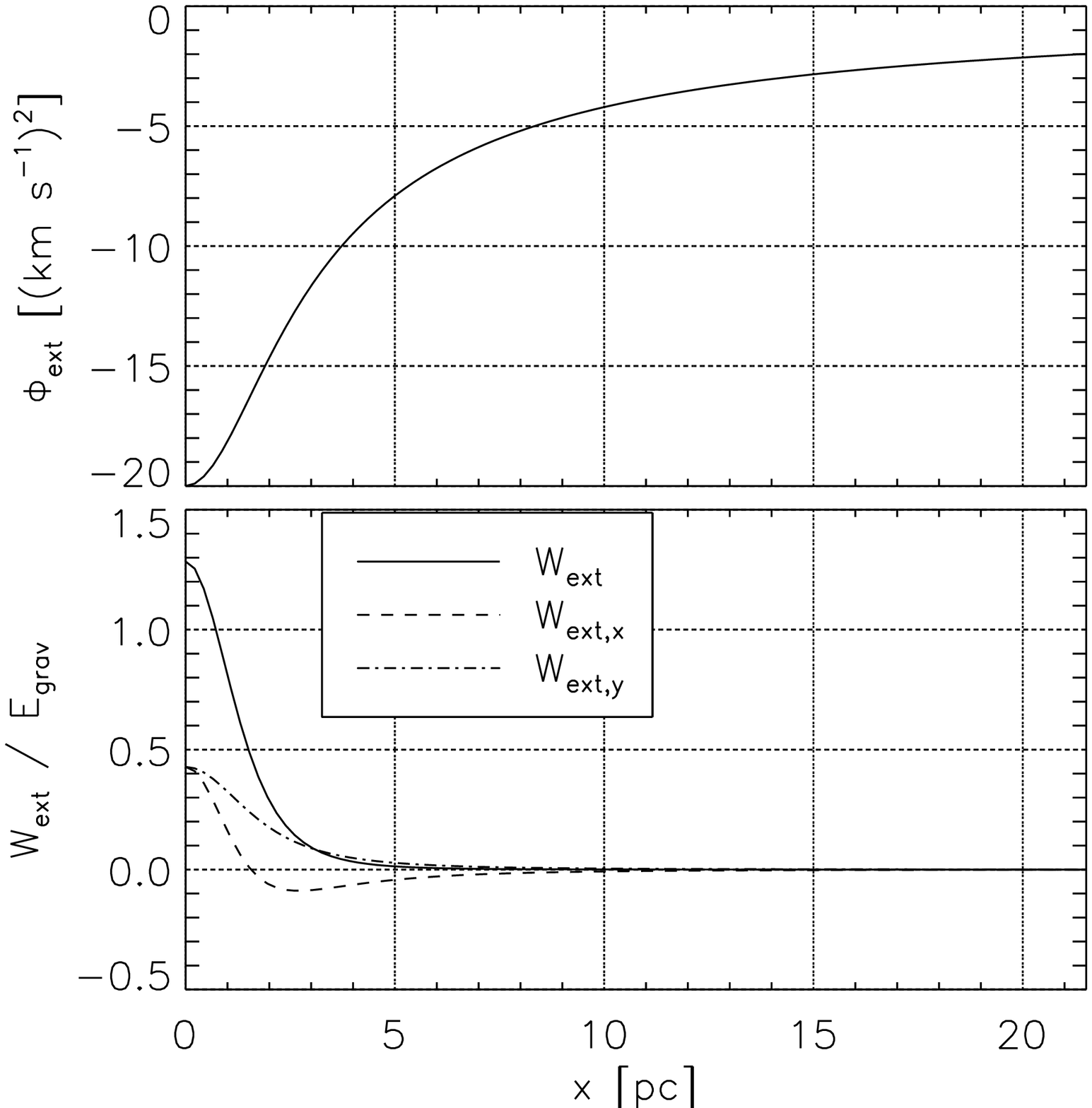}
  \caption{Upper panel: Plummer's potential as a function of the
  radius, which we take along the $x$ axis.  Lower panel:
  $\Wext/\Eg$ (solid line), $\Wextx/\Eg$ (dashed line) and
  $\Wexty/\Eg$ (dotted-dashed line) ratios for a $R_{\rm
  core}=0.5$~pc, $n=3500$~cm\alamenos 3 core embedded in a molecular
  cloud of \diezala 4 $\Msun$ following a \citet{Plummer1911}
  potential.  Note that the tidal contribution to the energetic
  content of the core ($\Wext$) are as important as the
  self-gravitational energy ($\Eg$) of the core.}
  \label{fig:plummer}
\end{figure}

Figure \ref{fig:plummer} shows both the potential (upper panel) and
the ratios $\Wext/\Eg$, $\Wextx/\Eg$, and $\Wexty/\Eg$ (lower panel)
as a function of distance to the center of the cloud, for our dense
core ($n=3500$~cm\alamenos 3, $R_{\rm core}=0.5$~pc) embedded in a
single Plummer potential.  As discussed in \S\ref{sec:tidal}, the
curvature of the potential defines whether $W$ is positive or
negative.  For instance, $\Wextx$ is negative whenever $\Phi_{\rm
ext}$ has upwards concavity, and positive wherever $\Phi_{\rm ext}$
has downward concavity.  On the other hand, $\Wexty$ is always
negative at every position along the $x$ axis because, along the
radius, $\Phi_{\rm ext}$ has upwards concavity in the azimuthal
direction.

More important, however, is to note in this figure that $\Wext$ grows
substantially when the core is located in the inner parts of the
molecular cloud.  In fact, it is larger (in absoluve value) than $\Eg$
at the central parts of the potential well, enhancing the importance
of accounting for $\Wext$ when analyzing the energy budget of dense
cores.

\begin{figure}
  \includegraphics[width=84mm]{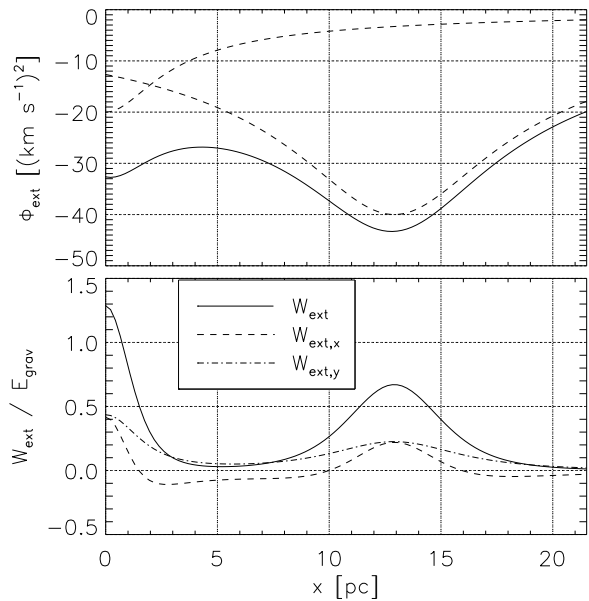}
  \caption{Same as fig.~\ref{fig:plummer}, but for a dense core
    embedded in a gravitational potential given by the superposition
    of two Plummer spheroids.  Dashed lines in the upper panel show
    the potential of the individual spheroids.  }
  \label{fig:plummer2}
\end{figure}

\begin{figure}
  \includegraphics[width=84mm]{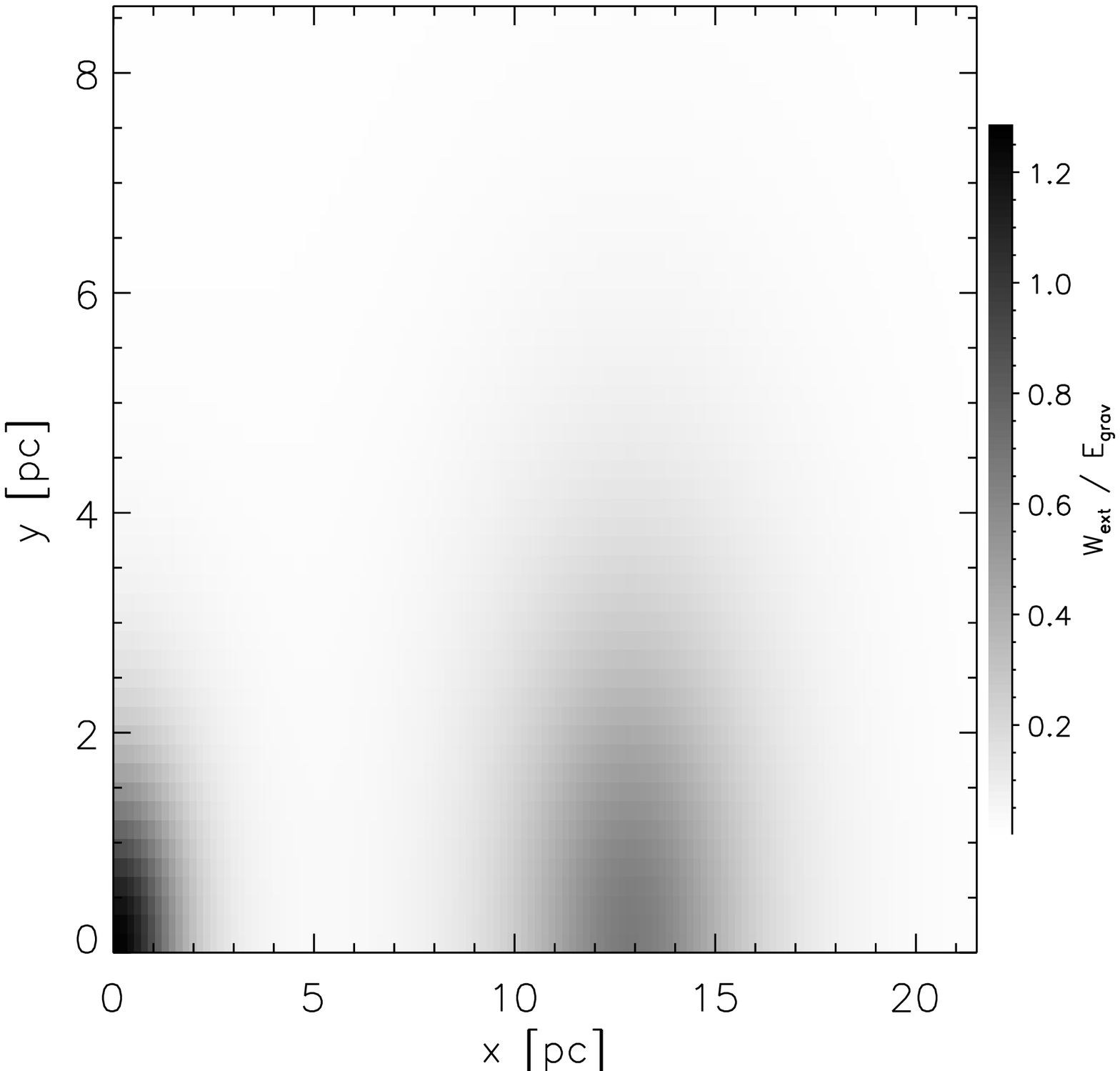}
  \caption{ Grayscale two-dimensional version of the lower panel of
    Fig.~\ref{fig:plummer2}.  Note that the scales in $y$ are expanded
    by a factor of 2.5.  The ratio $\Wext/Eg$ has cylindrical symmetry
    around the $x$ axis.}
  \label{fig:plummer2_gris}
\end{figure}

As a step towards taking a more realistic potential for a molecular
cloud, we show in Fig.~\ref{fig:plummer2} the potential (upper panel)
and the ratios $\Wext/\Eg$, $\Wextx/\Eg$, and $\Wexty/\Eg$ (lower
panel) for the same dense core embedded in the potential produced by
two Plummer spheroids, centered at different locations ($x=y=z0$~pc),
and $x=12.5$, $y=z=0$~pc), as a function of $x$.  In
Fig.~\ref{fig:plummer2_gris} we show a grayscale $x-y$ map of the
ratio $\Wext/\Eg$ on the $(x,y)$ plane.  As in the previous case, the
external field again contributes substantially to the energy balance
of the core, especially near the minimum of the potential well of the
cloud.  It should be noticed also that what defines the relative
importance of $\Wext$ is the curvature of the potential field, i.e.,
the width of the potential, and not only its depth (or height).  For
instance, the Plummer sphere centered at $x=12.5$, even though it is
more massive, is also more extended, producing a potential well with a
larger radius of curvature.  This translates into a lower value of
$\Wext$ at the position of the central region of that sphere, compared
to the contribution to $\Wext$ of the other Plummer sphere.

\begin{figure}
  \includegraphics[width=84mm]{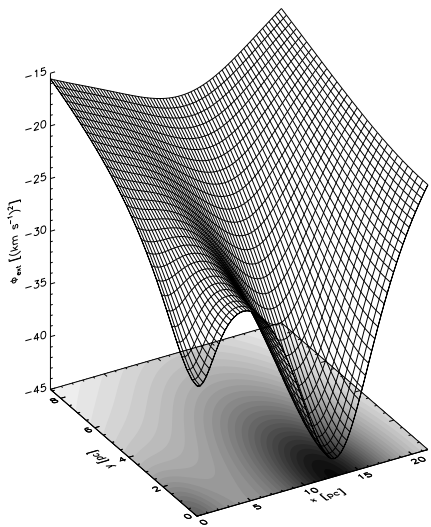}
  \caption{ Gravitational potential produced by the superposition of
    the two Plummer spheroids in the $x-y$ plane, both as a grayscale
    and surface representation.  Both minima occur at $y=0, z=0$,
    while at $x\sim 4$~pc occurs a saddle point.  A core located in
    this point will have strong compression if it is long enough along
    $y$, compared to its $x$-size, but strong disruption if it is long
    enough in $x$ compared to its $y$-size (note that the $x$ axis
    spans up to 20 units, while the $y$ axis spans only 8.)}
  \label{fig:sabana}
\end{figure}

{One may ask, in general terms, if this potential has places where it
can produce disruption or not.  To answer this question, in
Fig.~\ref{fig:sabana} we show the potential field of the last example
(two Plummer spheroids), both in grayscale and in a surface
representation, as a function of $x$ and $y$.  First of all, note that
the $x$ axis spans up to 20 units, while the $y$ axis spans only 8.
Second, note that the plot is symmetric with respect to the axes $x=0$
and $y=0$.  Third, note also that the dependence on $z$ (not shown
here) is the same as the dependence on $y$.

According to our discussion above, disruption occurs in regions with
predominant downwards concavity.  In our example (see
Fig.~\ref{fig:ejemplito}), the regions that are potentially disruptive
are (a) close to $x=4, y=0, z=0$~pc, and (b)regions at large $x$, $y$
and $z$. However, as can be seen in Fig.~\ref{fig:plummer2_gris},
$\Wext$ is always compressive ($\Wext/\Eg > 0$ everywhere).  The
reasons for this are that (a) At large $y$ (or $z$), the curvature
along $y$ is positive (upward concavity).  However, the curvature
along $x$ is negative, and with a smaller radius of curvature (see
Fig.~\ref{fig:sabana}, keeping in mind the different ranges spanned by
the axes).  Thus, the net effect of $\nabla \phiext$ over the cloud is
compressive.  (b) The local maximum at $x\sim 4, y=0, z=0$ is actually
a saddle point, and the concavity along $x$, although negative, has a
larger radius of curvature than the concavity along $y$, which is
positive (again, note that the $y$ axis is more compressed than the
$x$ axis).  Thus, in both (a) and (b) cases, the positive curvature
dominates, producing a positive value for $\Wext/\Eg$.  Then, this
potential produces compressions everywhere, at least for spherical
cores.}

\begin{figure}
  \includegraphics[width=84mm]{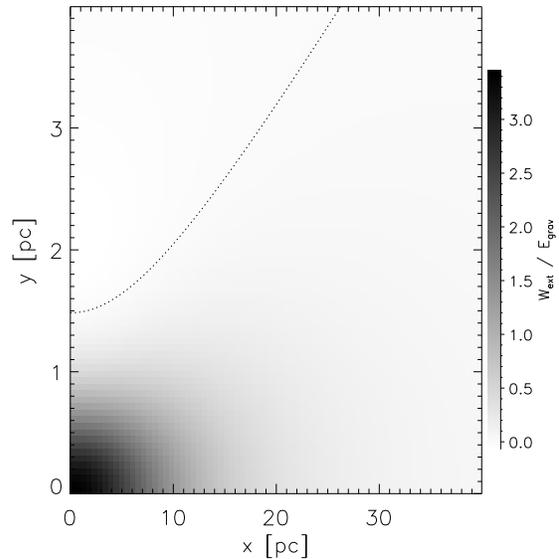}
  \caption{Grayscale of the ratio $\Wext/\Eg$ for a Plummer elongated
    spheroid, with aspect ratio 10.  The dotted line marks the
    $\Wext=0$ level (notice that the $y$ axis is expanded by a factor
    of 10).  Elongated structures seem to produce larger values of
    the ratio $\Wext/\Eg$.}
  \label{fig:LR10}
\end{figure}

Let us now consider an elongated molecular cloud, for which the
potential is given by a highly elongated Plummer spheroid that has an
aspect ratio of 10 (\S\ref{sec:core_setup}).  Fig.~\ref{fig:LR10}
shows the ratio $\Wext/\Eg$ in this case.  The dotted line marks the
place where the ratio is equal to zero.  We note that the maximum
contribution from $\Wext$ to the budget is now substantially larger
($\sim 3.5$ times the value of the gravitational energy $\Eg$).  This
is due to the fact that the density of a filament changes on short
scales along the direction perpendicular to its long axis, producing a
potential well with small radius of curvature in this direction.  {
Note also that the upper-left region, (above the dotted $\Wext/\Eg=0$
line), has negative values of the ratio $\Wext/\Eg$, i.e., positive
values of $\Wext$.  Although those values are small, the existence of
this region shows that there could be places where an isolated core,
in principle, might be disrupted by tidal forces from a larger cloud.
However, their small absolute values imply that, if such a core is in
energy equipartition between self-gravity and other forms of energy
(e.g., magnetic, turbulent, thermal), the tidal disruption will play a
negligible role, unless the core is substantially larger than the one
modeled here.}

{ In all the cases shown in the present section, the variation of
  the gravitational potential is soft enough that the external force
  $\nabla\Phi_{\rm ext}$ can be considered linear through the size of
  the core.  Thus, as in the case of the GMC embedded in a galactic
  gravitational field, $\Wext/\Eg \propto 1/\rho$,  and
  our results can be easily translated to cores with different
  densities.  The case of a core within a more realistic parent
  cloud (e.g., a filamentary GMC from numerical simulations)
  is deferred to a further contribution.  }

{ 

\subsection{General remarks}

In summary, for GMCs extended large enough in $z$ within a galactic
potential, the external term $\Wext$ is negative almost everywhere
because the compression along $z$ dominates.  However, for clouds
confined to the plane of the galaxy (for sizes smaller than 10~pc in
$z$), the galaxy may either compress or disrupt them, depending on the
relative orientation between the cloud and the direction of rotation
of the galaxy.  The spiral arms tend to produce compression, although
this also depends on the orientation and the size of the elongated
cloud.

In the case of the cores, where we have only considered the effects of
the gravitational potential of their parent cloud, the tidal energy
$\Wext$ of the structure embedded in a bigger potential can be
important close to the local minima, and acts in the same sense than
the gravitational energy: by compressing the cloud.  We have not found
any case in which $\Wext$ is positive and comparable to the
gravitational energy, suggesting that {\it the tidal energy of a GMC
over its cores helps in their confinement and collapse, but not in
their disruption.}

}



\section{Discussion and Conclusions}\label{sec:discussion}

Ranging from GMC sizes (from a few to several tens of parsecs), down to
prestellar core sizes ($\le 0.1$~pc), molecular clouds and their
substructure are often thought to be in virial equilibrium
\citep[e.g., ][ and references therein]{MO07}.  This state is inferred
from a comparison between their internal (kinetic, thermal, magnetic)
and gravitational energies.  In previous work, however, we have
emphasized the importance of accounting for the role of some
additional energy terms that should be considered, but that are
usually neglected when the virial theorem is used \citep[see ][]{BV97,
BVS99, BP06, Dib_etal07}.  In particular, \citet{BP06} argued that
there are at least six assumptions usually made in virial theorem
analyses that are frequently not fulfilled in the interstellar medium.
In the present work we have focused on the role of the net
external potential has on a density
structure (a cloud or a core). As a first attempt to
evaluate how realistic this assumption is, we have analyzed the
contribution of the mass external to the volume of integration to the
gravitational term $\Wext$, and compared it to the gravitational
energy, in two main cases: a GMC embedded within a spiral galaxy, and
a molecular cloud core within its parent molecular cloud.

{ 

Although our analysis does not exhaust the possibilities of cloud
and core shapes and configurations, the results
presented here suggest that by neglecting the influence of the
net potential external to a given core, observational estimates can
lead
to wrong estimations of its actual gravitational content.

We have found that the gravitational field of a Galaxy may have an
important influence on the energy budget of GMCs.  While the galactic
rotation tries to tear apart clouds that are elongated in the radial
direction, the
field along $z$ tend to compress the clouds.  The detailed
gravitational content of GMCs depends, thus, on their size, shape,
position and orientation within the galaxy.
This could be the case, for instance, of Maddalena's cloud, G216-2.5,
which exhibits no signs of massive star formation\citep[see,
  e.g.,][for the current status of star formation going on in
  G216-2.5]{Megeath_etal09}, in spite of its having a mass
  $\sim 6.6\times 10^5~\Msun$.

The situation for dense cores within simplified potentials for
molecular clouds is similar.  The gravitational field may tend to
disrupt the core in one direction, while compressing it in another.
The contribution of $\Wext$ depends strongly on the position of the
core inside the cloud, and on the details of the gravitational
potential of the later.  In all the cases shown here, however, this
contribution works in the same direction than the self-gravity, i.e.,
towards compressing the cloud.  Moreover, it can be of the order of
magnitude or even larger than the gravitational energy, specially in
elongated, filamentary clouds.  This is due to the fact that filaments
can have strong upwards concavity in the direction perpendicular to
their main axis.


An interesting result is that tidal disruption tends to be more
important than tidal compression for larger objects, while smaller
objects tend to suffer larger compressions.  The reason for this is
that there are no sources of gravitational repulsion, only
gravitational attraction.  Then, the radii of curvature when the
concavity is positive are smaller than when the concavity is negative.
This can be pictured by imagining the gravitational field as a tense
sheet supporting weights placed on it. On such arrangement, positive
(upwards) concavities at the position of the weights have small radii
of curvature, while negative (downwards) concavities always extend
over larger distances.
Thus, for a cloud of moderate size located where the external
gravitational field exhibits negative concavity, $\Wext$ will be
positive, but relatively small with respect to other forms of energy
(self-gravity, magnetic, turbulent, etc.).  However, if such a cloud
is located in a place where the curvature of the external field is
positive, its $\Wext$ will be important, since the field varies
strongly over smaller distances.

As a conclusion, the results shown here suggest that, on large scales,
i.e., GMCs within galaxies, the former can be farther from real
equipartition, either disrupted or compressed, depending on their
shape and orientation within the galaxy.  On smaller scales, it seems
that it is important to determine observationally the role of the
net external potential on local dense cores.

}

\section*{Acknowledgements}

We thank Paola D'Alessio and Lee Hartmann for a careful reading and
useful comments on this manuscript, and to Ian Bonnell, the referee,
for an encouraging and useful report.  This work was supported by
UNAM-PAPIIT grant numbers 110606 and IN119708 to JBP and BP,
respectively, and CONACYT grant numbers J50402-F, 50720, and U47366-F
to GCG, BP and EVS, respectively.  We have made extensive use of the
NASA-ADS database.  The calculations were performed on the cluster at
CRyA-UNAM acquired with grant 36571-E, and on the cluster Platform
4000 (KanBalam) at DGSCA, UNAM.

\label{lastpage}


\begin{thebibliography}{}

\bibitem[Allen \& Santill\'an(1991)]{Allen_Santillan91} Allen, C., \&
Santill\'an, A.\ 1991, Revista Mexicana de Astronom\'ia y
Astrof\'isica, 22, 255




  \bibitem[Ballesteros-Paredes(2006)]{BP06} Ballesteros-Paredes, J.\
  2006, \mnras, 372, 443

\bibitem[Ballesteros-Paredes \& Hartmann(2007)]{BH07}
Ballesteros-Paredes, J., \& Hartmann, L.\ 2007, Revista Mexicana de
Astronom\'ia y Astrof\'isica, 43, 123

\bibitem[Ballesteros-Paredes et al.(1999)]{BHV99} Ballesteros-Paredes,
J., Hartmann, L., \& V{\'a}zquez-Semadeni, E.\ 1999, \apj, 527, 285


\bibitem[Ballesteros-Paredes \& Mac Low(2002)]{BM02}
Ballesteros-Paredes, J., \& Mac Low, M.-M.\ 2002, \apj, 570, 734

\bibitem[Ballesteros-Paredes \& V\'azquez-Semadeni(1995)]{BV95}
Ballesteros-Paredes, J., \& V\'azquez-Semadeni, E.\ 1995, Revista
Mexicana de Astronom\'ia y Astrof\'isica, Conf. Series, 3, 105

\bibitem[Ballesteros-Paredes \& V\'azquez-Semadeni(1997)]{BV97}
Ballesteros-Paredes, J., \& V\'azquez-Semadeni, E.\ 1997, American
xInstitute of Physics Conference Series, 393, 81

\bibitem[Ballesteros-Paredes et~al.(1999a)]{BVS99}
{Ballesteros-Paredes}, J., {V\'azquez-Semadeni}, E., and {Scalo},
J. 1999a, \apj, 515, 286--303

\bibitem[Bertoldi \& McKee(1992)]{BM92} Bertoldi, F., \& McKee, C.~F.\
1992, \apj, 395, 140

\bibitem[Binney \& Tremaine(1987)]{BT87} Binney, J., \& Tremaine, S.\
1987, Princeton, NJ, Princeton University Press, 1987, 747 p.,

\bibitem[Blitz(1994)]{Blitz94} Blitz, L.\ 1994, The Cold Universe, 99




 \bibitem[Brice\~no et al.(1997)]{Briceno_etal97} Brice\~no, C.,
Hartmann, L.~W., Stauffer, J.~R., Gagne, M., Stern, R.~A., \&
Caillault, J.-P.\ 1997, \aj, 113, 740


\bibitem[Chandrasekhar \& Fermi(1953)]{Chandra_Fermi53} Chandrasekhar,
S., \& Fermi, E.\ 1953, \apj, 118, 116

\bibitem[Dame et al.(2001)]{Dame_etal01} Dame, T.~M., Hartmann, D., \&
Thaddeus, P.\ 2001, \apj, 547, 792

\bibitem[de Jong et al.(1980)]{deJong_etal80} de Jong, T., Boland, W.,
\& Dalgarno, A.\ 1980, \aap, 91, 68

\bibitem[Dib et al.(2007)]{Dib_etal07} Dib, S., et al. 2006, \apj, submitted

\bibitem[Downes et al.(1996)]{Downes_etal96} Downes, D., Reynaud, D., 
Solomon, P.~M., \& Radford, S.~J.~E.\ 1996, \apj, 461, 186




\bibitem[Falgarone et al.(1991)]{Falgarone_etal91} Falgarone, E.,
Phillips, T.~G., \& Walker, C.~K.\ 1991, \apj, 378, 186




\bibitem[Gieles et al.(2006)]{Gieles_etal06} Gieles, M., Portegies 
Zwart, S.~F., Baumgardt, H., Athanassoula, E., Lamers, H.~J.~G.~L.~M., 
Sipior, M., \& Leenaarts, J.\ 2006, \mnras, 371, 793 

\bibitem[Goldreich \& Kwan(1974)]{Goldreich_Kwan74} Goldreich, P., \&
Kwan, J.\ 1974, \apj, 189, 441

\bibitem[G{\'o}mez et al.(2007)]{Gomez_etal07} G{\'o}mez, G.~C.,
V{\'a}zquez-Semadeni, E., Shadmehri, M., \& Ballesteros-Paredes, J.\
2007, \apj, 669, 1042



\bibitem[Hartmann et al.(2001)]{HBB01} Hartmann, L.,
Ballesteros-Paredes, J., \& Bergin, E.~A.\ 2001, \apj, 562, 852



\bibitem[Heiles \& Troland (2005)]{Hei05} {Heiles}, C., \& {Troland},
T. H. 2005, \apj, 624, 773

\bibitem[Helfer et al.(2003)]{Helfer_etal03} Helfer, T.~T., Thornley, 
M.~D., Regan, M.~W., Wong, T., Sheth, K., Vogel, S.~N., Blitz, L., \&
Bock, D.~C.-J.\ 2003, \apjs, 145, 259

\bibitem[Herbig(1978)]{Herbig78} Herbig, G.~H.\ 1978, Problems 
of physics and evolution of the universe, p.~171 - 179, 180 - 188, 171 

\bibitem[Herbig et al.(1986)]{Herbig_etal86} Herbig, G.~H., Vrba, 
F.~J., \& Rydgren, A.~E.\ 1986, \aj, 91, 575 

\bibitem[Hunter \& Fleck(1982)]{Hunter_Fleck82} Hunter, J.~H., Jr., \&
Fleck, R.~C., Jr.\ 1982, \apj, 256, 505





\bibitem[Kegel(1989)]{Kegel89} Kegel, W.~H.\ 1989, \aap, 225, 517


\bibitem[Klessen et~al.(2000)]{KHM00} {Klessen}, R.~S., {Heitsch}, F.,
and {Mac Low}, M.-M. (2000) \apj, 535, 887--906



\bibitem[Larson(1981)]{Larson81} {Larson}, R.~B. 1981, \mnras 194,
09--826



\bibitem[Loinard et al.(1999)]{Loinard_etal99} Loinard, L., Dame, 
T.~M., Heyer, M.~H., Lequeux, J., \& Thaddeus, P.\ 1999, \aap, 351,
1087

\bibitem[Loinard et al.(1996)]{Loinard_etal96} Loinard, L., Dame, 
T.~M., Koper, E., Lequeux, J., Thaddeus, P., \& Young, J.~S.\ 1996,
\apjl, 469, L101

 

\bibitem[Maloney(1988)]{Maloney88} Maloney, P.\ 1988, \apj, 334, 
761 





\bibitem[McKee \& Ostriker(2007)]{MO07} McKee, C.~F., \& Ostriker,
E.~C.\ 2007, \araa, 45, 565


\bibitem[McKee and Zweibel(1992)]{MZ92} {McKee}, C.~F. and {Zweibel},
E.~G. 1992, \apj 399, 551--562

\bibitem[Megeath et al.(2009)]{Megeath_etal09} Megeath, S.T.,
  Allgaier, E., Young, E., Allen, T., Pipher, J.L., \& Wilson,
  T.L. 2009. \apj, submitted


\bibitem[Miyamoto \& Nagai(1975)]{Miyamoto_Nagai75} Miyamoto, M., \&
  Nagai, R.\ 1975, \pasj, 27, 533


\bibitem[Myers \& Goodman(1988a)]{MG88a} Myers, P.~C., \& Goodman,
A.~A.\ 1988a, \apjl, 326, L27

\bibitem[Myers \& Goodman(1988b)]{MG88b} Myers, P.~C., \& Goodman,
A.~A.\ 1988b, \apj, 329, 392







\bibitem[Pichardo et al.(2003)]{Pichardo_etal03} Pichardo, B., Martos,
M., Moreno, E., \& Espresate, J.\ 2003, \apj, 582, 230

\bibitem[Plummer(1911)]{Plummer1911} Plummer, H.~C.\ 1911, \mnras, 
71, 460 


\bibitem[Roberts et al.(1979)]{Roberts_Huntley_vanAlbada79} Roberts,
W.~W., Jr., Huntley, J.~M., \& van Albada, G.~D.\ 1979, \apj, 233, 67



\bibitem[Scalo(1990)]{Scalo90} Scalo, J.\ 1990, ASSL Vol.~162:
Physical Processes in Fragmentation and Star Formation, 151

\bibitem[Schmidt(1956)]{Schmidt56} Schmidt, M.\ 1956, \bain, 13, 15

\bibitem[Shadmehri et al.(2002)]{Shadmehri_etal02} Shadmehri, M.,
V{\'a}zquez-Semadeni, E., \& Ballesteros-Paredes, J.\ 2002, ASP
Conf.~Ser.~276: Seeing Through the Dust: The Detection of HI and the
Exploration of the ISM in Galaxies, 276, 190


\bibitem[Shu et al.(1987)]{SAL87} Shu, F.~H., Adams, F.~C., \& Lizano,
S.\ 1987, \araa, 25, 23

\bibitem[Spitzer(1978)]{Spitzer78} Spitzer, L.\ 1978.  Physical
processes in the Interstellar Medium. New York Wiley-Interscience


\bibitem[Tilley \& Pudritz(2004)]{TP04} Tilley, D.~A., \& Pudritz,
R.~E.\ 2004, \mnras, 353, 769

\bibitem[Troland \& Crutcher(2008)]{Troland_Crutcher08} Troland,
T.~H., \& Crutcher, R.~M.\ 2008, ArXiv e-prints, 802, arXiv:0802.2253

\bibitem[Vázquez-Semadeni et al.(1997)]{VBR97} Vázquez-Semadeni, E.,
Ballesteros-Paredes, J., \& Rodriguez, L.~F.\ 1997, \apj, 474, 292




\bibitem[V{\'a}zquez-Semadeni et al.(2007)]{VS_etal07}
V{\'a}zquez-Semadeni, E., G{\'o}mez, G.~C., Jappsen, A.~K.,
Ballesteros-Paredes, J., Gonz{\'a}lez, R.~F., \& Klessen, R.~S.\ 2007,
\apj, 657, 870

\bibitem[V\'azquez-Semadeni et al.(2008)]{VS_etal08}
V\'azquez-Semadeni, E., Gonz\'alez, R.F., Ballesteros-Paredes, J.,
Gazol, A., \& Kim, J. 2008. \mnras, submitted






\bibitem[Wilson et al.(1970)]{Wilson_etal70} Wilson, R.~W., Jefferts,
K.~B., \& Penzias, A.~A.\ 1970, \apjl, 161, L43


\bibitem[Young \& Scoville(1991)]{Young_Scoville91} Young, J.~S., \& 
Scoville, N.~Z.\ 1991, \araa, 29, 581 

\bibitem[Zuckerman \& Evans(1974)]{Zuckerman_Evans74} Zuckerman, B.,
\& Evans, N.~J., II 1974, \apjl, 192, L149

\end{thebibliography}
 \end{document}